\documentclass[letterpaper]{article} 
\usepackage{aaai23}  
\usepackage{times}  
\usepackage{helvet}  
\usepackage{courier}  
\usepackage[hyphens]{url}  
\usepackage{graphicx} 
\usepackage{natbib}  
\usepackage{caption} 
\usepackage{algorithm}
\usepackage{algorithmic}

\usepackage{newfloat}
\usepackage{listings}
\DeclareCaptionStyle{ruled}{labelfont=normalfont,labelsep=colon,strut=off} 
\floatstyle{ruled}
\newfloat{listing}{tb}{lst}{}
\floatname{listing}{Listing}
\title{The Design and Organization of Educational Competitions with Anonymous and Real-Time Leaderboards in Academic and Industrial Settings}
\author{
    Serdar Kad{\i}o\u{g}lu\textsuperscript{\rm 1, \rm 2},
    Bernard Kleynhans\textsuperscript{\rm 1}
}
\affiliations{
    \textsuperscript{\rm 1 } AI Center of Excellence, Fidelity Investments, Boston, USA \\
    \textsuperscript{\rm 2 } Department of Computer Science, Brown University, Providence, USA
}

\usepackage{bibentry}

\begin{document}

\maketitle

\begin{abstract}
The goal of this paper is to share our experience in designing and organizing educational competitions with anonymous and (near) real-time leaderboards in both academic and industrial settings. While such competitions serve as a great educational tool and provide participants with hands-on experience, they require significant planning, technical setup, and administration from organizers. In this paper, we first outline several important areas including team registration, data access, submission systems, rules and conditions that organizers should consider when planning such events. We then present a high-level system design that can support (near) real-time evaluation of submissions to power anonymous leaderboards and provide immediate feedback for participants. Finally, we share our experience applying this abstract system in academic and industrial settings. We hope the set of guidelines and the high-level system design proposed here help others in their organization of similar educational events. 
\end{abstract}

\section{Introduction}
The field of Artificial Intelligence (AI) has seen tremendous progress over the last decade. Many problems that seemed to be completely out of reach can now be handled routinely, e.g., in autonomous game-play, natural-language processing, and computer vision. One of the driving forces in improving the state-of-the-art has been well-designed competitions that provide researchers and practitioners with an opportunity to reach broader audiences and to objectively evaluate the performance of their algorithms. Today, there exist several well-established competitive events aimed at tracking our progress in the field. 

In academia, notable events include ACM RecSys Challenge~\cite{said2016short}, CVPR Computer Vision Challenge~\cite{demir2018deepglobe,lomonaco2020cvpr}, ICAPS Automated Planning and Scheduling Challenge~\cite{vallati20152014} and International SAT Solver Competition~\cite{jarvisalo2012international}. These events have been exceedingly successful. For instance, in computer vision, the ImageNet Large Scale Visual Recognition Challenge (ILSVRC)~\cite{russakovsky2015imagenet} for object detection and classification over millions of images with hundreds of categories sparked the deep learning revolution. 

In industry, companies are required to constantly deliver new products and services to remain competitive and offer value to their customers. Many of these companies have adopted various strategies such as running hackathons, not only to shorten the product development cycle but also to maximize the contribution of their talent for innovation. For instance, the Netflix \$1M prize increased the accuracy of the Netflix recommendation system by more than 5\%~\cite{bennett2007netflix}. Hackathons have been adopted not only in corporations of all sizes but also in education~\cite{Nandi2016HackathonsAA,Decker2015UnderstandingAI}. Traditionally, AI Competitions are aimed at improving the state-of-the-art on established benchmarks. Similarly, Industry Competitions crowd-source solutions for immediate business problems. In parallel, Educational Competitions, the topic of this paper, are geared toward certain learning outcomes and engagement.

\begin{figure*}[t]
	\begin{center}
\includegraphics[width=140mm,keepaspectratio]{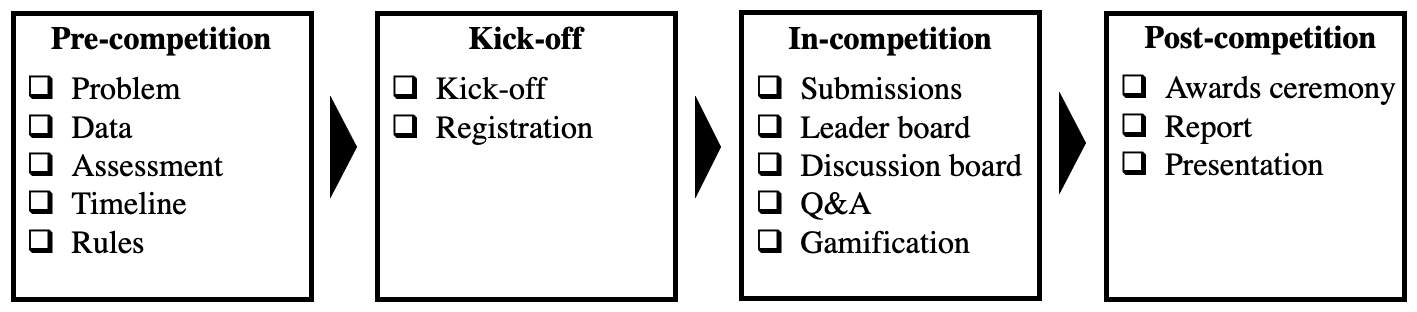}
\caption{Components of a competition across the chronological stages before, during, and after the competition.}
	\label{fig:components}
	\end{center}
\end{figure*}

In this paper, we would like to share our experience in designing and conducting competitions with a particular focus on the educational context in both academic and industrial settings. Overall, we make three main contributions: 

\begin{enumerate}
    \item \textbf{Competition Planning:} We start with an outline of a comprehensive set of important considerations and guidelines to help plan and organize such educational competitions. 
    \item \textbf{System Design:} We then present a high-level system design to support the necessary back-end architecture to conduct competitions. Our system design is tool-agnostic and platform-independent, e.g., Kaggle~\cite{Kaggle}, Codalab~\cite{Codalab}, that might not be viable options due to technical overhead, data privacy, and intellectual property. An important design consideration is to support near real-time evaluation of results to provide immediate feedback for participants via anonymous leaderboards. 
    \item \textbf{Experience Reports:} Finally, we share our experience from applying this abstraction in practical scenarios within academic and industrial settings at Brown University and Fidelity Investments, respectively. 
\end{enumerate} 

We hope that our outline for competition planning, system design, and experience reports serve as a starting point for researchers and practitioners in their efforts to organize similar educational events.

\section{Competition Planning}
\label{sec:competition}

At a high level, we split competition planning into four chronological stages each of which is associated with a set of components as shown in Figure~\ref{fig:components} and detailed next. 

\subsection{Pre-competition}

\textbf{Problem definition:} The abstract problem and the learning objectives of the competition should be defined clearly. It might be an existing problem that is well-known to the participants. From an educational perspective, the problem should allow solutions of varying complexity; challenging to motivated participants, and welcoming for beginners. In our experience, framing the abstract problem in the real-world context to narrate its impact (e.g., efficient use of resources, social good, and etc.) is an important motivator. In industry, business stakeholders can step in with their domain expertise to provide background and can actively collaborate and learn from AI experts in a low-pressure setting.

\smallskip

\noindent \textbf{Data:} The abstract problem definition is realized with specific instances of the problem. The data should exhibit sufficient richness to enable a range of approaches and the data protocol (e.g., distinguishing train and test in machine learning context) should be defined apriori. Separate holdout data for the final evaluation can help avoid over-fitting. Common data formats, e.g., csv, should be used to lower the entry barrier. Ideally, data structures and naming conventions should be consistent with the domain standard, and including metadata describing each dataset is useful. 

\smallskip

\noindent \textbf{Assessment:} Metric(s) used to evaluate submissions should be precisely defined using a formula or code. Ideally, it is inherited from a common metric from the domain. Providing an implementation of the metric helps avoid misinterpretations. If multiple metrics are used, careful consideration of how to combine the different metrics is required. 

\smallskip

\noindent \textbf{Timeline:} The timeline and the sequence of events should be decided ahead of the competition. A registration period before the competition kick-off helps participants get familiar with the problem, data and submission process. The duration should be sufficiently long and account for the difficulty of the problem, data preparation required, and training and solving time for baseline approaches. We strongly suggest setting a preliminary deadline early in the competition that requires an initial submission from participants to continue. In the classroom setting, this encourages students to become familiar with the problem and iterate on their solutions.

\smallskip

\noindent \textbf{Rules and conditions:} Governance is required for several criteria including eligibility, team size, modifications and restrictions (e.g., cannot participate in more than one team), collaboration and discussion policy, data access, and usage rights during and after the competition, conditions on additional data usage, submission constraints (e.g., a daily limit), deliverables during and at the end of the competition (e.g., sharing reproducible source code to be eligible for the prizes), timelines, and the official time zone of the event. 

\subsection{Kick-off}

\noindent \textbf{Kick-off event:} An official event (or announcement) to introduce the problem, spike interest, point out available resources, provide registration instructions and contact details.

\smallskip

\noindent \textbf{Registration:} Registration should require little effort from participants and enable automation for organizers. This step plays a crucial role in our System Design as explained later.

\subsection{In-competition}

\noindent \textbf{Leaderboard:} The leaderboard tracks the scores,  number of submissions, and other relevant metrics for each team. Frequently updated leaderboards help participants receive prompt feedback, learn, and remain engaged. Anonymity retains privacy while still providing benefits from feedback.

\smallskip

\noindent \textbf{Discussion board:} Discussion boards allow participants to communicate with each other and organizers. This addresses frequently asked questions where communication is broadcasted to everyone. The rules and conditions should cover the discussion policy (e.g., on sharing direct answers or source code and the code of conduct).

\smallskip

\noindent \textbf{Q\&A sessions:} Following the kick-off event, facilitating online Q\&A sessions provide participants an opportunity to ask technical questions about the problem and data once they are more familiar with the setup.

\smallskip

\noindent \textbf{Gamification:} Gamificiation has been shown to improve learning in education~\cite{dicheva2015gamification}. A universal theme such as Olympics can highlight collaborative and competitive spirit. Similarly, awarding badges at pre-defined milestones (e.g., first submission, first to pass a baseline, most creative team name) boosts  participation. Another option is to introduce an in-competition twist (e.g., additional dataset, different evaluation metric, solution complexity). This pushes participants to adapt and be creative.  
It is advised to recognize the leaderboard status prior to the change.

\begin{figure*}[t]
	\begin{center}
\includegraphics[width=120mm,keepaspectratio]{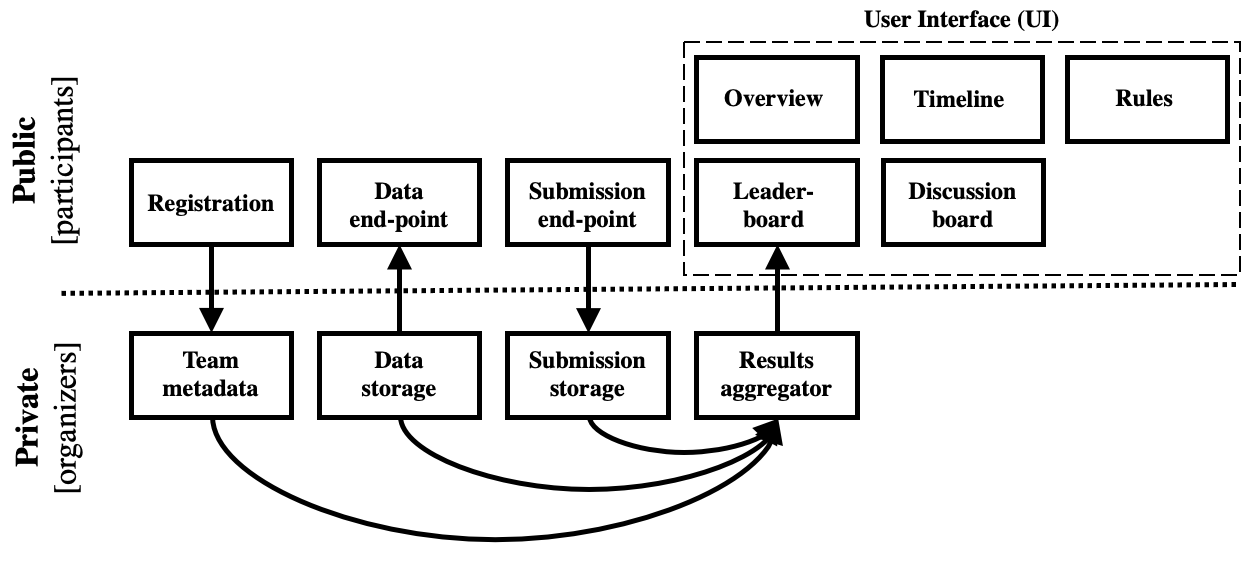}
\caption{High-level system design with public and private components.}
	\label{fig:system}
	\end{center}
\end{figure*}

\newpage

\subsection{Post-competition}

\noindent \textbf{Awards ceremony:} This is the ceremony to announce and celebrate competition winners and prizes. This is also an opportunity to be creative in recognizing every participant's effort and be inclusive beyond top-ranked teams. Organizers should award behavior and hard work that helps the community (e.g., most answers on the discussion board, teams at the K-12 level, solo participants, etc.).

\smallskip

\noindent \textbf{Report \& presentation:} 
Beyond the top performers, all participants are encouraged to summarize their approach in a report. Best solutions can further have an opportunity to present their approach to maximize learning for everyone. Reports that remain accessible post-competition foster future learnings even for non-participants at a later time. 

\section{System Design}
\label{sec:system}


Next, we present a high-level system design of the necessary functionality to host educational competitions. The system allows near real-time evaluation of results that can be presented in an anonymous leaderboard. Online platforms such as Kaggle~\cite{Kaggle} and Codalab~\cite{Codalab} offer similar functionality and should also be considered. However, the type of problems and evaluation protocols that are supported by these platforms is limited. Additionally, for data privacy and intellectual property reasons these platforms are often not viable options in an industrial setting. 

\smallskip

Figure~\ref{fig:system} presents the overall system design. The system is composed of two main parts; the public front-end and the private back-end. 

\smallskip

From the participants' perspective, the public front-end provides access to registration, a data end-point, a submission end-point, a leaderboard, a discussion board, and information and resources related to the competition. Parts or all of the front-end components can be exposed in a single user interface (UI) as depicted in the encircling box in Figure~\ref{fig:system}. 

\smallskip
From the organizers' perspective, the private back-end is composed of team metadata for bookkeeping captured by the registration process, data storage to share competition instances with participants using a public data end-point, submission storage to cache solutions flowing from the public submission end-point, and finally, a compute source to aggregate results in private and update the (anonymous) leaderboard in public.

\smallskip

The registration process plays a crucial role in linking the public and private components together. As part of the registration process, teams declare their participant information anonymous leaderboard name, and a declaration to comply with the rules and terms of the competition. This information is captured in team metadata to be utilized later. 

\smallskip

Notice that our system design leaves the choice of specific tools open. There are many tools that can be used for each of the system components. 

\smallskip

In Section~\ref{sec:competition}, we outlined the components of a competition, and then, in Section~\ref{sec:system}, we presented an abstract system design to support the necessary functionality. Next, in Section~\ref{sec:brown} and Section~\ref{sec:fidelity}, we bring these building blocks together and provide specific instantiations in practice based on competitions conducted in academia and in industry.

\section{Experience Report: Brown University}
\label{sec:brown}

The Foundations of Prescriptive Analytics course, CSCI-2951O\footnote{\label{courseweb}\url{https://cs.brown.edu/courses/csci2951-o/}}, has been taught at the Department of Computer Science at Brown University every year since 2016. On average, {\raise.17ex\hbox{$\scriptstyle\mathtt{\sim}$}}25 students enroll the course, with double the enrollment number is waitlisted due capacity constraints. The main learning objective of this graduate-level course is to provide students with a comprehensive overview of the theory and practice of optimization technology. A wide variety of state-of-the-art techniques are studied: Boolean Satisfiability~\cite{biere2009handbook}, Constraint Programming~\cite{rossi2006handbook}, Integer/Linear Programming~\cite{wolsey1999integer}, and Local Search \& Meta-Heuristics~\cite{hoos2004stochastic}. 

\begin{table*}[t]
\centering
\renewcommand{\arraystretch}{1.3}
\begin{tabular}{|c|c|c|c|}
\hline
Project & Paradigm                    & Business Domain          & Application          \\ \hline \hline
I       & Boolean Satisfiability      & Automative Industry      & Mass Customization   \\ \hline
II      & Constraint Programming      & Human Capital Management & Workforce Scheduling \\ \hline
III     & Linear Programming          & Supply Chain Management  & Facility Location    \\ \hline
IV      & Integer Programming         & Healthcare Analytics     & Testcase Diagnosis   \\ \hline
V       & Local Search \& Meta-heuristics & Logistics            & Transportation       \\ \hline
\end{tabular}
\caption{Overview of course projects in CS2951 at Brown University each conducted as an educational competition.}
\label{tab:projects}
\end{table*}

\smallskip
As shown in Table~\ref{tab:projects}, CS2951o is a hands-on course with projects that are designed to cover business-relevant applications. Each paradigm is coupled with a project to solve challenging benchmark instances from its respective domain. The projects ask students to either use off-the-shelf general-purpose constraint solvers to model and solve the problem, implement custom algorithms from scratch, or combine the two approaches together to create hybrid solutions.

\smallskip

CS2951o fits neatly with the competition and system design presented here. Each project is conducted as an educational competition over {\raise.17ex\hbox{$\scriptstyle\mathtt{\sim}$}}3 weeks with an anonymous and (near) real-time leaderboard. Students submit their solutions on the given benchmark instances and receive immediate feedback on how it ranks relative to other submissions.

\subsection{System Design}

\begin{figure*}[!h]
	\begin{center}
\includegraphics[width=100mm,keepaspectratio]{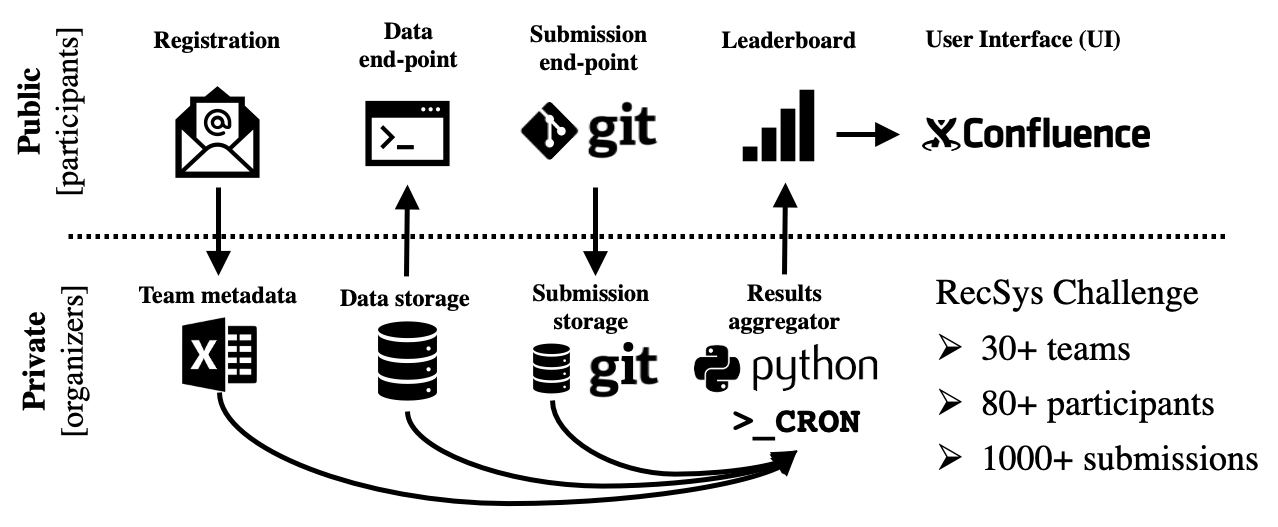}
\caption{System design for the RecSys competition @ Fidelity Investments}
	\label{fig:system_fidelity}
	\end{center}
 \vspace{-0.1cm}
\end{figure*}

Let us introduce the particular instantiation of our abstract system from Figure~\ref{fig:system} when running projects in CS2951 and highlight specific tools. We use the same setup and architecture for each project.

\smallskip

\noindent \textbf{UI front-end:} We relied on Piazza and the course website as the main user interface and communication medium. For each project, the problem definition, data description, evaluation metric, timeline, rules, resources were shared on Piazza. The leaderboard was hosted on the course website. 

\smallskip

\noindent \textbf{Registration:} In the course setting, the registration refers to collecting anonymous team names for the leaderboard. However, having multiple projects within the semester leads to a dilemma between collaboration and privacy. 

\smallskip

\noindent \textbf{Collaboration \& anonymity:} On one hand we would like students to have the option to collaborate (in teams of two), and on the other hand, we want to maintain student anonymity in subsequent projects. Our system design supports this conflicting requirement as follows. In the first week, we assign a special project (Project - 0) where each student must submit five anonymous tokens, one for each project. When collaborating, students reveal their anonymous token to each other. When teams change in the following projects, students still remain anonymous with the freedom to work with a different partner. As a side-benefit, this simple project gives students a chance to get familiar with the project structure and the submission system early on. Also, notice how our system design allows team members to work in parallel and make independent submissions when working on alternative approaches.

\smallskip

\noindent \textbf{Data end-point:} We again use Piazza to share benchmark instances and support code. For datasets, we select benchmark instances from the research literature that react differently to different algorithmic approaches that are covered in lectures. As such, a new submission can improve on some instances while worsening others. Every semester, we often encounter a niche solution, with subpar overall results but stellar performance on specific instances. This leads students re-consider the definition of the \textit{best solution} and experience first-hand that there is no silver bullet for solving hard combinatorial problems.

\smallskip

\noindent \textbf{Submission end-point:} Brown University offers an in-house submission tool. It is a command-line utility used in most courses that copies artifacts into the course directory within the specific project and student folders. In CS2951o, we ask that each submission is accompanied by a results log file with evaluation metric(s) on each benchmark instance for the project at hand. In terms of metrics, common choices are solution quality, optimality/feasibility status, and runtime. Students start with partial submissions that leave some instances unsolved, and then, gradually improve.

\smallskip

\noindent \textbf{Results aggregator:} In the back-end, there is a cron job that walks through each submission folder every 5 seconds. The result log files are processed into an aggregated csv file. Notice that our update frequency does not leave enough time to re-run each submission. Instead, we rely on the results provided to refresh the leaderboard quickly. In parallel, there is an overnight cron job that compiles and runs student submissions to verify the results submitted. 

\smallskip

\noindent \textbf{Leaderboard:} In the front-end, we use a D3.js visualization for the leaderboard on course website. It tracks the aggregated csv file making results available with each submission. This is by far the most highlighted aspect of CS2951o in course reviews. Anecdotally, every semester we receive student feedback praising the leaderboard and mentioning how it kept them engaged and pushed their approaches further. The students suggest that more courses adopt this approach.

\section{Experience Report: Fidelity Investments}
\label{sec:fidelity}

The AI Center of Excellence at Fidelity Investments organized a Recommender Systems (RecSys) competition for employees.
The main learning objective of the competition was to provide employees hands-on experience with recommender systems and to generate interest in this area.

\smallskip

\noindent \textbf{Problem:} The competition was based on a classical content recommendation problem where for a given set of users (subscribers) and a given set of items (articles) one has to determine the best $k$ items to show to each user. Specifically, participants had to determine which 10 articles, ranked by their relevance should be recommended to each subscriber. The problem captures the common 

\smallskip

\noindent \textbf{Data:} Historical interaction data indicating which articles have been clicked or not-clicked by each subscriber in the past were made available to participants. To enable other researchers and practitoners the anonymous data and publicly available articles shared with the community~\cite{DBLP:journals/corr/abs-2307-04996}\footnote{https://github.com/fidelity/mab2rec/tree/main/data}. Additionally, we also provided attributes for each article and subscriber that could be used as features in a recommendation algorithm. The data was formatted to have a similar structure to well-known Recommender System benchmark datasets, e.g., MovieLens~\cite{harper2015movielens}.

\smallskip

\noindent \textbf{Resources:} 
Apart from data, we shared a variety of learning resources and tools to help  participants get up to speed with recommender systems. Among other alternatives, we covered open-source software developed within Fidelity to increase familiarity and internal adoption. These open-source libraries are accessible to everyone and mainly include: 

\begin{enumerate}
 \item Feature selection and generation: \textsc{Selective\footnote{https://github.com/fidelity/selective}}~\cite{kadiouglu2021optimized,ijcai_active_learning}, \textsc{Seq2Pat\footnote{https://github.com/fidelity/seq2pat}}~\cite{seq2pat,ai_magazine,KDF_dpm, DPM_frontiers}, and \textsc{TextWiser\footnote{https://github.com/fidelity/textwiser}}~\cite{textwiser2021}
    \item Recommendation models: 
    \textsc{Mab2Rec\footnote{https://github.com/fidelity/mab2rec}}~\cite{mab2rec}, and \textsc{MABWiser\footnote{https://github.com/fidelity/mabwiser}}~\cite{mabwiser_ictai,mabwiser_ijait,DBLP:journals/tmlr/KilitciogluK22}

    \item Recommendation performance and fairness evaluation: \textsc{Jurity\footnote{https://github.com/fidelity/jurity}}~\cite{DBLP:conf/icmla/MichalskyK21,DBLP:conf/cikm/ChengKK22}
\end{enumerate}

\smallskip

\noindent \textbf{Assessment:} Submissions were evaluated on a test dataset using Mean Average Precision (MAP), a commonly used recommender system evaluation metric for ranking tasks. We provided a formula for the metric made accesible to  participants via Jurity library.

\smallskip

\noindent \textbf{Timeline:} Participants had one week to register, six weeks to make submissions and one week to submit a report. We required participants to make at least one submission by the end of the second week for continued participation. After three weeks we also introduced a change to the evaluation protocol to only consider items (articles) that were present in the test data. At this point we froze the leaderboard for the first part of the competition and created a new leaderboard to rank submissions using the modified evaluation. This twist renewed interest in the competition and leveled the playing field by penalizing algorithms that did not generalize well.

\smallskip

The competition took place in the midst of the  pandemic with employees working from home and in hybrid setups. The event was successful in terms of the engagement and learning goals. In total, we had 30+ teams participate with 80+ individuals. The majority of the group had background in machine learning and data science. During the 6-week event more than 1,000 submissions were made with substantial daily traffic on the leaderboard throughout the competition. At the end, the competition page generated more traffic than the \textit{total page visits} from popular internal company pages. Importantly, the competition provided an opportunity for employees working in different parts of the company to collaborate and socialize while working remotely.

\subsection{System Design}
As before, let us share the particular instantiation of our abstract system shown in Figure~\ref{fig:system} when running RecSys. Let us note that we do not advocate any particular tool, we rather present a high-level architecture that can accommodate different instantiations based on available tools. With our design, most suitable tools can be used interchangably. 

\smallskip

Figure~\ref{fig:system_fidelity} highlights the specific tools used in this particular case. Notice that the tools are commonly used by practitioners and are either open-source or standard in industrial settings. Our implementation is configurable and can serve as a reference for future organizers. 

\smallskip

\noindent \textbf{UI front-end:} We used Atlassian Confluence to create a competition site as the  front-end for participants. It included relevant information about the competition such as problem definition, data description, evaluation metrics, timeline, rules, learning resources, and contact details for any issues or questions. It also hosted a discussion board and an anonymous leaderboard. 

\smallskip

\noindent \textbf{Registration:} Participants registered using a registration button on the competition site that opened an email with a pre-filled subject line and body that had to be completed and sent to the organizers. The team information was manually entered into a spreadsheet with team metadata and registration was confirmed by the organizers via email.

\smallskip

\noindent \textbf{Data end-point:} The prepared datasets (csv files) and data dictionary was copied to a shared server that all registered participants were given access to. The registration confirmation email sent to participants included instructions on how to copy the data using a simple \texttt{scp} command that participants could execute from command-line.

\smallskip

\noindent \textbf{Submission end-point:} For submissions we utilized GitLab as a Git repository manager. A private repository was created for each team to which results could be pushed. Each push constituted a submission. 

\smallskip

\noindent \textbf{Results aggregator:} A cron job ran an aggregator script every minute, which would pull results from all the Git repositories, evaluate, and combine the scores for each team into a single aggregated csv file. The MAP evaluation of all submissions is fast enough to support the availability of results every minute for the cron job.

\smallskip

\noindent \textbf{Leaderboard:} We created a simple Confluence page with a leaderboard table and a bar chart to visually compare results among the teams and show the total number of submissions for each team. The table and the chart point to the aggregated csv file to display the latest results. The cron job updates the results every minute, as such, the leaderboard provides prompt feedback to participants. Participants were highly engaged with the leaderboard, submitting solutions often and comparing their results. The leaderboard also helped to fix issues that are not related to performance, e.g., incorrect submission format. Notice, the leaderboard allows participants to remain anonymous by displaying team names provided in the registration stage.

\section{Conclusion}

In this paper, we first outlined a comprehensive set of important considerations and guidelines to help plan and organize competitions. We then presented a high-level system design to support the necessary back-end architecture to conduct such competitions. Finally, we shared our experience from practical scenarios within academic and industrial settings. At a first glance, our guidelines for planning and our high-level system design for infrastructure might appear obvious. However, the successful organization of a competition demands careful design and thorough planning. Failure to do so requires additional effort from organizers \textit{during} the competition resulting in an undesirable experience for participants. Ultimately, we can only justify the time and resources investment from multiple parties with careful planning in advance. 
With this in mind, we take a step toward bringing different components together. We welcome feedback from the community and hope that others can utilize this as a starting point to organize even better educational events to train AI practitioners in a hands-on and collaborative setting.

\bibliography{aaai23}

\end{document}